\PassOptionsToPackage{algo2e,ruled,vlined}{algorithm2e}
\documentclass{article}
\usepackage{spconf,amsmath,graphicx,hyperref}

\usepackage{caption}
\usepackage{cite}
\usepackage{amsmath,amssymb,amsfonts}
\usepackage[linesnumbered,ruled,vlined]{algorithm2e}

\usepackage{graphicx}
\usepackage{textcomp}
\usepackage{xcolor}
\usepackage{amsmath,amssymb}
\usepackage{algorithm} 
\usepackage{balance}
\usepackage{algpseudocode}
\usepackage{bm}
\newcommand{\y}{\mathbf{y}}

\newcommand{\br}{\mathbf{r}}

\newcommand{\x}{\mathbf{x}}
\newcommand{\by}{\mathbf{y}}

\newcommand{\bGamma}{\boldsymbol{}{\Gamma}}

\newcommand{\cX}{\mathcal{X}}
\newcommand{\cY}{\mathcal{Y}}
\newcommand{\cN}{\mathcal{N}}
\newcommand{\bR}{\mathbb{R}}
\def\minwrt[#1]{\underset{#1}{\text{minimize }}}
\def\maxwrt[#1]{\underset{#1}{\mathrm{maximize }}}
\def\argminwrt[#1]{\underset{#1}{\text{arg min }}}

\colorlet{LightBlue}{blue!60!green}

\title{Acoustic image source interpolation with Optimal transport barycenter}
\name{Yuyang Liu\textsuperscript{+}\textsuperscript{$\ddagger$} Rumeshika Pallewela\textsuperscript{+}\textsuperscript{$\ddagger$}, Jesper Brunnström\textsuperscript{*}, Isabel Haasler\textsuperscript{*},  Filip Elvander\textsuperscript{+}\thanks{This research was supported in part by the Research Council of Finland
(decision number 362787).\\Authors 1 and 2 have made equal contributions to this paper.}}
\address{\textsuperscript{+}Dept. of Information and Communications Engineering, Aalto University, Finland\\\textsuperscript{*} Dept. of Information Technology, Uppsala University, Sweden }
   
\begin{document}
\fontsize{9.3}{11.4}\selectfont
%
\maketitle
\begin{abstract}
\small
Room impulse responses can be estimated via the image source model (ISM) using the image source point cloud (ISPC) of a physical source. 
However, because the source movement changes the ISPC, estimating the ISPC at a new source position typically requires repeated acoustic measurements.
We propose an optimal transport (OT) barycenter framework to interpolate the ISPC of a new source location from ISPCs of known sources. The method jointly estimates image-source associations and the ISPC at the new location. The OT ground cost exploits the
property 
that the image sources undergo the same displacement as their physical sources. This approach is realized for both grid-based and support-free configurations. 
The support-free method addresses the resulting nonconvex joint estimation problem by alternating between identifying image-source associations across the ISPCs and refining the target image-source locations. This enables the interpolation of ISPCs without repeated measurements, facilitating efficient and flexible room-acoustic modeling.
\end{abstract}

\vspace{-2mm}
\section{Introduction}
\vspace{-2mm}
\label{Intro}

A Room impulse response (RIR) characterize the acoustic properties of enclosed environments. Accurate estimation of RIRs is essential for a broad range of room acoustic applications, including indoor noise cancellation \cite{KoyamaNoiseCancel,WaterschootDereverb}, speech enhancement \cite{2010SpeechEnhancement}, and auralization \cite{tervo2013spatialAuralization}. A measured RIR is commonly modeled as being comprised of direct sound, early reflections, late diffuse reverberant field, and measurement noise \cite{kuttruff2016room}. In the early stage of sound propagation, the direct sound and the earliest boundary reflections dominate the RIR. Under the assumptions of geometrical acoustics, these reflections can be modeled by the image source model (ISM) \cite{allen1979image}, in which each sequence of specular reflections is represented by a virtual source obtained by mirroring the physical source across the corresponding reflecting boundary. 

\begin{figure}[t]
    \centering
    \includegraphics[width=\columnwidth]{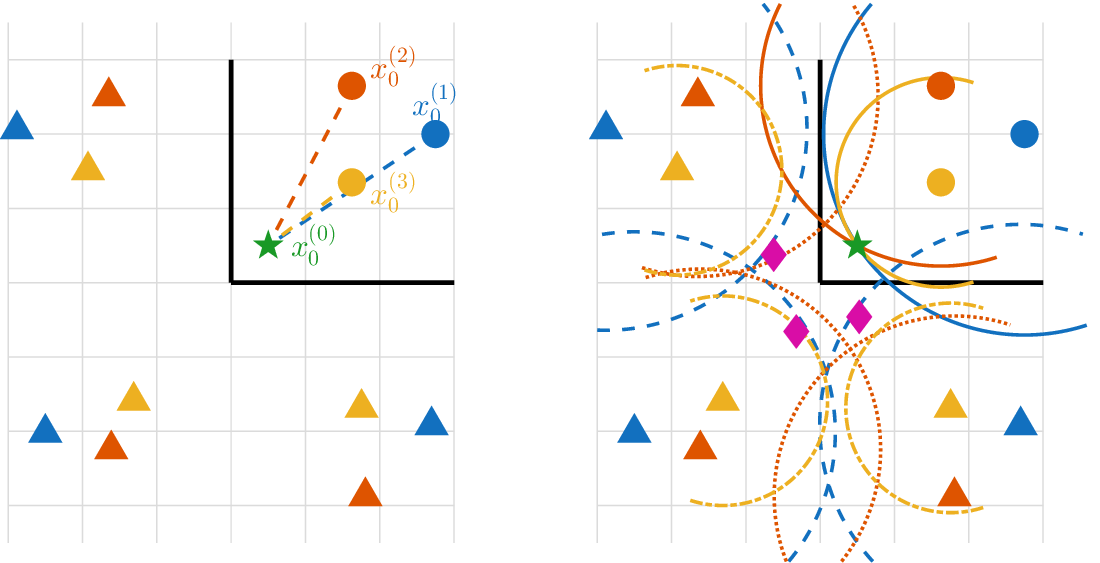}
    
    \caption{\small 
    Illustration of estimating a new ISPC from three known source configurations. Left (Geometry): known physical sources (colored circles) and their corresponding image sources (triangles with matching colors); the target physical source is the green star. Right: For each configuration, circles are centered at the known image sources with radii equal to the displacement from the corresponding physical source to the target; the resulting loci determine the interpolated image-source locations (magenta diamonds). 
    }
    \label{fig:illustration}
\end{figure}

Under the ISM, early RIRs at any receiver can be estimated once the image source point cloud (ISPC) of the physical source is known~\cite{allen1979image}. Consequently, prior work has focused on recovering the ISPC for stationary source positions from multi-receiver measurements~\cite{sprunck2022gridless,Ribeiro2012roomModeling,puomio2021locating,carlo2021dechorate}. These methods estimate the ISPC at one or more known source locations and then synthesize RIRs at receiver positions. However, when the source moves, the ISPC changes, typically requiring additional measurements and a fresh ISPC estimation, which is a process not straightforward in the current literature. 
Inferring the room geometry provides a solution as discussed in~\cite{dokmanic2013acoustic,sprunck2025fully,kelley2024rir}. However, in practice, geometric errors tend to compound for higher-order reflections, and estimates based on a limited set of source positions can produce biased or incomplete room models. Motivated by these limitations, we aim to estimate the ISPC at a new source position directly from ISPCs obtained at previously measured positions; without explicitly recovering the room geometry. A key \textit{observation} is that a small displacement of the physical source induces an equal displacement of each corresponding image source, although its direction remains unknown in the absence of room-geometry information. This displacement invariant motivates the geometric penalty used to interpolate the ISPC at a new source position from previously estimated ISPCs. Note that image-source locations may exhibit positional perturbations due to boundary-induced scattering and sampling quantization error~\cite{ochmann2004complex,cox2006tutorial}.

 In general, the realization is of two parts: 
\textit{pairing same order image sources (association)} and \textit{localizing new ISPC}. Previous work solved the association problem using validation gating and nearest neighbor rules
\cite{4323129}, probabilistic data association \cite{BarShalom1975TrackingIA}, clustering
\cite{7953335}, and multidimensional assignment based on feasible measurement combinations \cite{8455616}, \cite{7077851}. Herein, we propose an optimal transport (OT) barycenter approach to interpolate the image sources for a moved physical source. Recent studies have applied OT and OT barycenters in room acoustics to estimate RIRs or image sources of receivers~\cite{geldert2023interpolation,sundstrom2024optimal,pallewela2025room}. For the ISPC interpolation task, we construct the OT cost from the displacement invariance. The framework operates naturally on a fixed grid, and we also provide a grid-free implementation through a free-support OT barycenter~\cite{lindheim2023simple}. 

\vspace{-2mm}
\section{Signal Model}
\vspace{-2mm}
\label{SigMod}
Consider a room $\Omega \subset \bR^d$, for $d = 2$ or $d=3$,  with an impulse point source located at $\x_0 \in \Omega$ and a microphone located at $\br\in \Omega$.
Then, the RIR is the solution of the wall-bounded inhomogeneous wave equation \cite{pierce2019acoustics}, with the early part of the signal being composed of the arrival of direct sound and  reflections from the wall. The ISM regards these reflections as sound emitted from point sources in free field: ideally, these image sources at locations are obtained by recursively mirroring the original source across the reflecting walls~\cite{allen1979image}. However, in practice, due to the wave scattering and sampling quantization error, the \textit{observed} or \textit{effective} image source $\x_i\in\mathbb{R}^d\setminus\Omega$ may be offset from the ideal mirrored positions by a position perturbation.

Therefore, the ideal mirrored image sources are regarded as unperturbed images sources and the observed ones are perturbed image sources. Then, the early RIR can be approximated using the physical source $\x_0$ and its image source point cloud (ISPC) $\cX = \{\x_i\}_{i=1}^{N}$ and 
\begin{equation*}
    h_e(\br,t) = \sum_{i} \frac{w_i}{\lVert \x_i-\br \rVert_2}\delta(t-\frac{\lVert \x_i-\br \rVert_2}{c}),
\end{equation*}
where $N$ equals the number of image sources and $w_i$ is the weight of the $i$-th image source.
%
%
%
%
Thus, RIRs at any receiver $\br\in\Omega$ can be estimated once the ISPC of a specific known source $\x_0$ is given. In contrast to a moved receiver, knowing ISPC of a \textit{moved source} is not a straightforward problem as the entire ISPC changes with the source. We address it with the observation that the reflected image sources have equal displacement to the physical ones.

\subsection{ISPC of a new source} \label{sig_mod2}
Consider the ISPC of a known source $\x_0\in\Omega$, denoted by $\cX=\{\x_i\}_{i\in\mathcal I}$, where $\mathcal I=\{1,..,i,..,N\}$ is the set of latent reflection labels and $N$ denotes the number of image sources.
Suppose the physical source is moved from $\x_0$ to a new position $\by_0 \in \Omega$ with displacement $\rho=\lVert\x_0-\by_0\rVert_2$, where $\y_0$ has ISPC $\cY=\{\y_j\}_{j=1}^{N}$. $\cY$ has the same number of image sources as $\cX$ and corresponds to the same set of latent reflections. If ISPCs are unperturbed, all image sources are moved by the same distance $\rho$, and
%
%
%
%
%
there exists a permutation $\pi:i\to j$ such as 
\begin{equation}
    \lVert \x_{\pi(j)}-\by_j\rVert_2 = \rho.
    \label{eq:distance_preservation}
\end{equation}








We extend to $K$ known sources located at $\x_0^{(k)}$, $k=1,\dots,K$, each associated with an ISPC $\cX^{(k)}$. All ISPCs contain the same number of reflections and correspond to the same set of latent reflections. If all ISPCs are unperturbed, each $\cX^{(k)}$ has the displacement preserving relation with $\cY$ given by Eq.~\eqref{eq:distance_preservation}, with respect to a distance $\rho^{(k)}$ and a permutation $\pi^{(k)}$. These relations constrain the locations of image sources and enable interpolation of $\cY$ once $\pi^{(k)}$ is given. Consequently, this leads to the necessities to find the permutation $\pi^{(k)}$ and the ISPC $\cY$. However, in practice, the observed ISPCs are perturbed, which introduces deviations from the ideal distance-preserving relations. To address the tasks, we adopt an iterative scheme that alternates between associating the objects with displacement invariance properties of the given ISPCs and updating the interpolated ISPC $\cY$ with the given associations. The association step can be formulated as an optimal transport (OT) barycenter problem, where the ground-cost construction determines the interpolation geometry and, consequently, the interpolated point.





\subsection{Optimal association barycenter} \label{OT}
OT is a mathematical framework for computing the minimum transportation cost required to transform one probability measure into another \cite{peyre2020computationaloptimaltransport,villani2009optimal}.  In OT framework, the ISPC is formed as an equally weighted discrete probability measure,
\begin{equation*}
    \mu^{(k)}(\x) = \sum_{i =1,\dots,N} \frac{1}{N} \delta(\x-\x_i^{(k)}), \quad\!\! \nu (\y) = \sum_{j =1,\dots,N} \frac{1}{N} \delta(\y-\by_j),
\end{equation*}
where $\delta(\cdot)$ denotes the Dirac delta measure and 
$N$ is the number of image sources.
%
%
%
With equal weights, the interpolated ISPC is obtained as
\begin{equation}
\begin{aligned}
\min_{\substack{\cY,\,
\bm P^{(k)}\in\mathbb R_+^{N\times N}}}
&\quad
\sum_{k=1}^{K}
\left\langle
\bm C^{(k)}(\cY),\bm P^{(k)}
\right\rangle_{\mathrm F} \\
\mathrm{s.t.}
&\quad
\bm P^{(k)}\bm 1_N=N^{-1}\bm 1_N,\quad
\bm P^{(k)\top}\bm 1_N=N^{-1}\bm 1_N ,
\end{aligned}
\label{Corr_OT}
\end{equation}
where $\langle\bm A,\bm B\rangle_{\mathrm F}
=\operatorname{tr}(\bm A^\top\bm B)$ is the Frobenius inner product,
$\bm P^{(k)}$ is the $k$th transport plan, and
\begin{equation}
\left[\bm C^{(k)}(\cY)\right]_{ij}
=
\left(
\left\|\x_i^{(k)}-\by_j\right\|_2-\rho^{(k)}
\right)^2 .
\label{cost}
\end{equation}
In the absence of perturbations, the correct permutation $\pi_k$
satisfies
$[{\bm C}^{(k)}({\cY})]_{\pi^{(k)}(j),j}=0$. Thus, the association permutation transport yield zero total
cost.


\section{Proposed Method}
\vspace{-2mm}
For interpolating the new ISPC with $K$ ISPCs $\cX^{(k)}$, with $k=1,\cdots,K$, we consider two estimators, depending on whether a sufficiently fine spatial grid is computationally practical. When such a grid is practical, the target ISPC is estimated over the prescribed support, leading to a convex linear program. Otherwise, the target support is estimated directly in $\mathbb{R}^{d}$ by alternating between OT-based data association and support interpolation. Accordingly, the OT cost is constructed using the perturbed image-source positions and the candidate support as defined in \eqref{cost}.

\subsection{Grid-based association barycenter}
\label{ssec:grid_barycenter}
%

%
Let $\mathcal G=\{\bm g_q\}_{q=1}^{Q}\subset\mathbb R^d$ be a grid and
$\mathcal B_Q=\{\bm b\geq\bm0:\bm1_Q^\top\bm b=1\}$. 
The grid-based association barycenter solves:
\begin{equation}
\begin{aligned}
 \min_{\substack{
\bm b\in\mathcal B_Q\\
 \bm P^{(k)}\in\mathbb R_+^{N\times Q}}}\quad&
 \sum_{k=1}^{K}
 \langle\bm C^{(k)}(\mathcal G),\bm P^{(k)}\rangle_{\mathrm F}\\[-1mm]
 \mathrm{s.t.}\quad&
 \bm P^{(k)}\bm1_Q=N^{-1}\bm1_N,\quad
 (\bm P^{(k)})^\top\bm1_N=\bm b,\quad\forall k .
\end{aligned}
\label{eq:grid_barycenter_problem}
\end{equation}
Here, $\bm b$ contains the grid masses and $\bm P^{(k)}$ is the
$k$th transport plan. Since
$\mathcal G$ is fixed, Eq.
\eqref{eq:grid_barycenter_problem} is a linear program with the standard structure of a fixed support discrete Wasserstein barycenter \cite{lin2020fixed} as its solution $\widehat{\bm b}$ defines the estimated measure
$\widehat\nu_{\mathcal G}=\sum_q\widehat b_q\delta_{\bm g_q}$ and selecting the $N$ grid points with the largest masses yields the estimated ISPC $\widehat{\mathcal Y}_{\mathcal G}$.
The estimator is convex, with accuracy limited by the grid resolution.

\subsection{Free-support association barycenter}

\label{ssec:free_barycenter}

To reduce the computational cost associated with a sufficiently fine grid, the support $\mathcal{Y}$ is instead optimized directly in the continuous spatial domain. Unlike Eq.~\eqref{eq:grid_barycenter_problem},
the joint problem is nonconvex: the objective
$\sum_k\langle\bm C^{(k)}(\mathcal Y),\bm P^{(k)}\rangle_{\mathrm F}$  couples the unknown support and transport plans, and its cost entries are generally
nonconvex in $\mathcal Y$. We therefore alternate between association
and interpolation updates~\cite{cuturi2014fast}.

\subsubsection{Step 1: OT-based data association}

Given the current support
$\mathcal Y$, we construct
$\bm C^{(k)}(\mathcal Y)$ using
\eqref{cost} and solve the fixed-support version of
Eq.~\eqref{Corr_OT} for 
$\{\bm P^{(k)}\}_{k=1}^{K}$.
This is a linear program, where
$P_{ij}^{(k)}$ is the mass assigned from $\x_i^{(k)}$ to
$\y_j$ and therefore represents their soft association.

\subsubsection{Step 2: Support interpolation}

For fixed transport plans, minimizing the barycenter objective over
the target support $\mathcal Y$ decomposes into independent problems
for its support points.
Accordingly, each point is updated as
\begin{equation}
\by_j
=
\arg\min_{\by\in\mathbb R^d}
\sum_{k=1}^{K}\sum_{i=1}^{N}
P_{ij}^{(k)}
\Bigl(
    \|\bm{x}_{i}^{(k)}-\by\|_2-\rho^{(k)}
\Bigr)^2 .
\label{eq:support_interpolation}
\end{equation}

Thus, $\by_j$ is the weighted ground-cost mean of the image
sources softly assigned to it. In the algorithm, the numerical solution
of \eqref{eq:support_interpolation} is denoted by
$\operatorname{Intp}(\cdot)$ and is computed using the eigenvalue-based
weighted squared-range approximation of~\cite{larsson2025single}.
The algorithm is applied separately using the initialization strategies
considered in Section~\ref{ssec:initialization}. In each run, the association and interpolation steps are alternated, with the support points updated independently during the interpolation step. The iterations terminate when either the root mean square (RMS) coordinate change $\eta$ between successive support estimates satisfies $\eta \leq \tau$, where $\tau$ is a prescribed tolerance, or the maximum number of iterations $L_{\max}$ is reached. In the definition of $\eta$ given in Algorithm~\ref{alg:ot_barycenter}, $d$ and $N$ denote the spatial dimension and the number of image sources in the target ISPC, respectively.


\subsection{Initialization}
\label{ssec:initialization}

The free-support estimator is nonconvex in $\mathcal{Y}$, and its solution may depend on the initial support. To ensure a fair comparison, we evaluate the following initialization strategies using the same number of candidate points, determined by the \emph{All-PC circular} construction with $M$ points sampled per circular locus:
\begin{itemize}
    \item \textbf{Uniform:} candidate points are sampled uniformly over a region containing the available ISPCs,
    \item \textbf{All-PC circular:} candidate points are sampled on the
    radius-$\rho^{(k)}$ loci centered at image sources from all $K$ available
    ISPCs,
    \item \textbf{One-PC circular:} candidate points are sampled on the radius-$\rho^{(k)}$ loci associated with one available ISPC.
\end{itemize}

\begin{algorithm2e}[t]
\small
\SetAlgoVlined
\caption{Alternating free-support ISPC interpolation
}
\label{alg:ot_barycenter}
\KwIn{$\mathcal S=\bigcup_{k=1}^{K}\mathcal X^{(k)}$,
initial support $\mathcal Y_{\mathrm{init}}=\{\by_j\}_{j=1}^{N}$},
$\{\rho^{(k)}\}_{k=1}^{K}$, $N$, tolerance $\tau$, and $L_{\max}$
\KwOut{Estimated target ISPC $\widehat{\mathcal Y}$}
Initialize $\mathcal Y\leftarrow\mathcal Y_{\mathrm{init}}$\;
\For{$\ell=0,\ldots,L_{\max}-1$}{
       Store $\by_{j,\mathrm{old}}\leftarrow\by_j$\;     
    \tcp{Step 1: OT data association}
    \For{$k=1,\ldots,K$}{
        
         Construct $\bm C^{(k)}(\mathcal Y$)
        using~\eqref{cost}\;
        
        Solve the fixed-support problem~\eqref{Corr_OT}
        for $\bm P^{(k)}$\;
        
    }

    \tcp{Step 2: Support interpolation}
    \For{$j=1,\ldots,N$}{
    
        
        $\y_j
        \leftarrow
        \operatorname{Intp}\!\left(
            \left\{
            \bigl(
                \bm x_i^{(k)},
                \rho^{(k)},
                P_{ij}^{(k)}
            \bigr)
            \right\}_{k,i}
        \right)$\;
    }
\tcp{Compute the support change - $\eta$ }
$\eta\leftarrow 
\left( \frac{1}{dN} \sum_{j=1}^{N} \|\by_j-\by_{j,\mathrm{old}}\|_2^2 \right)^{1/2}$\\\
    \If{$\eta<\tau$}{
        \textbf{break}\;
    }
}
\Return{$\widehat{\mathcal{Y}} {\leftarrow\mathcal{Y}}$}\;
\end{algorithm2e}
\section{Numerical Experiments and results}

\begin{figure*}
    \centering
    \includegraphics[
        width=\textwidth,
        height=0.16\textheight
    ]{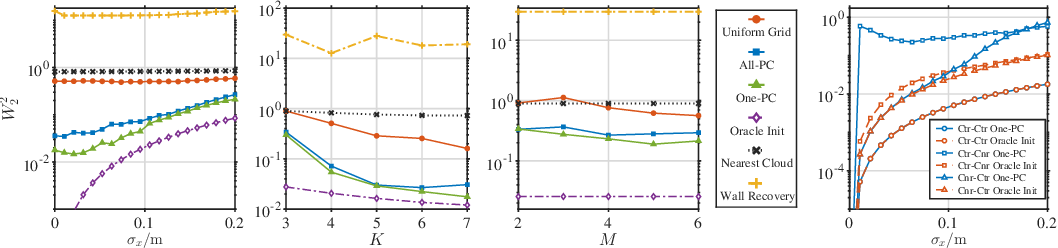}
     \caption{ 
     \small
     Mean squared Wasserstein distance, $W_2^2$, between the estimated and ground-truth ISPC. From left to right, the performance is shown as a function of: (a) the image-source perturbation standard deviation $\sigma_x$ for the different initialization strategies, oracle initialization which is initialized with unperturbed points; (b) the number of known physical sources $K$; (c) the number of sampled points $M$ for each ring in ALL-PC; and (d) the perturbation level for the center--center (Ctr--Ctr), center--corner (Ctr--Cnr), and corner--center (Cnr--Ctr) source configurations.}
    \label{fig:W2}
\end{figure*}

\begin{figure}[t]
    \centering
    \includegraphics[width=\columnwidth]{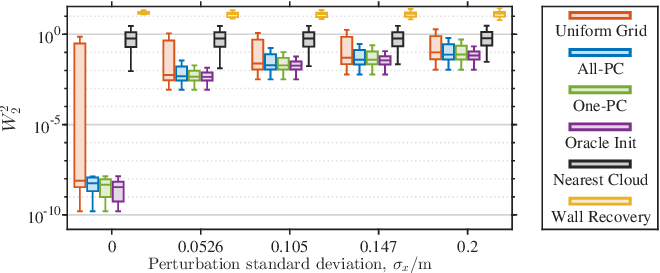}
\caption{\small Squared Wasserstein distance $W_2^2$ between the estimated and ground-truth ISPC for different initialization $\sigma_x$. The boxplots 
summarize 400 Monte Carlo trials}    \label{fig:boxplot}
\end{figure}

In this section, the performance of the proposed method for estimating new ISPC is evaluated using simulated 2D data. In the simulations, the ISPCs are generated in a $3\times4$\,m rectangular room while the room geometry is assumed unknown to the interpolation methods.
For each physical source, the corresponding ISPC contains $N$ image sources associated with reflections up to the second order, generated according to \cite{allen1979image}. To model uncertainties caused by wall scattering and sampling quantization errors, each image source is perturbed by an independent Gaussian displacement 
$\boldsymbol{\Delta}_i^{(k)}\sim \cN(0,\sigma_x^2\mathbf{I})$
, where $\mathbf{I}$ is the $2\times2$ identity matrix. The $K$ known physical sources are placed on a uniformly spaced circular array. The proposed method is evaluated using the three initialization strategies described in Section~\ref{ssec:initialization}. In addition, we include an oracle aided benchmark, denoted
\emph{Oracle Init}, in which the algorithm is initialized with the unperturbed point cloud. 
Furthermore, as noted in section \ref{ssec:initialization}, $M$ is defined as the number of candidate points sampled on each ring in the \emph{All PC} method.
We compare the proposed method against two baselines, one of which assumes that the interpolated ISPC is the same to the one of nearest physical source to $\y_0$, whereas the other reconstructs the room geometry following \cite{dokmanic2013acoustic}, with echo labeling performed using the nearest-neighbor principle. Each method is evaluated using the average squared 2-Wasserstein distance between the estimated ISPC $\hat{\cY}$ and the unperturbed ISPC, 
defined as 
\footnote{This Wasserstein distance is
used only as an evaluation metric. Unlike the proposed association
cost in Eq.~\eqref{cost}, it uses the squared Euclidean distance
between estimated and reference image-source locations.}
\begin{equation*}
    W_2^2 = \frac{1}{N} \min_{\bGamma}\left\langle C_e, \bGamma \right\rangle_{F} \  \ \text{s.t.} \ \bGamma 1_{N} =N^{-1}\bm1_N,\ \bGamma^{\top} 1_N = N^{-1}\bm1_N,
\end{equation*} 
where the $(i,j)$-th entry of $C_e$ is the squared Euclidean distance between the $i$-th point in $\hat{\cY}$ and the $j$-th point in the unperturbed ISPC. Three parametric studies are first conducted by varying the perturbation standard deviation $\sigma_x$, the number of known sources $K$, and the search-space resolution parameter $M$. The known sources $\x_0^{(k)}$ are uniformly distributed on a circle of radius 0.5 m centered in the room, while the new source is randomly placed. The fixed default parameters are $\sigma_x=0.1m$, $K=4$, and $M=4$. The results, averaged over 400 Monte Carlo trials, are shown in the first three subfigures of Fig.~\ref{fig:W2}. The \emph{One-PC} initialization yields the lowest estimation error
among the practical initialization strategies. Furthermore, the performance gap to \emph{Oracle Init} counterpart indicates that data association errors degrade performance. The $\sigma_x$ curve indicates that the uniform initialization is comparatively flat, and this is due to the large initialization error that dominates the perturbation errors. To further examine this behavior, Fig.~\ref{fig:boxplot} shows the distributions over Monte Carlo trials, illustrating the variability caused by random source positions. The results also show the proposed methods improve with more known ISPCs but it's insensitive to the initialization grid resolution. The 
Fig.~\ref{fig:W2} (d) demonstrates the geometrical accuracy of the method to the spatial configuration. Here, \emph{Ctr} and \emph{Cor} denote the room centre and a position near its lower-left corner, respectively. In each configuration label, the first term (\emph{Ctr} or \emph{Cor}) specifies the centre of the circular arrangement of known sources $\{\x_0^{(k)}\}_{k=1}^{K}$, while the second specifies the target-source position $\by_0$. For example, \emph{Ctr--Cor} places the known-source arrangement at the room centre and $\by_0$ near the corner. The \emph{Ctr--Ctr} configuration achieves performance comparable to that of the oracle method. In contrast,
the \emph{Ctr--Cor} and \emph{Cor--Ctr} errors increase sharply with perturbation and also
demonstrate that interchanging the locations of $\x^{(k)}_0$ and $\by_0$ does not lead to equivalent performance. This behavior may arise because, when physical sources lie near room boundaries, their image sources become denser and more interleaved, making the interpolated support more sensitive to perturbations and the correct associations harder to recover. This experiment suggests that, unlike in the first three studies, data-association errors are not the only dominant limitation in these geometrically difficult cases but the perturbations are strongly amplified by the spatial configuration. 

\vspace{-2mm}
\section{Conclusion}
\vspace{-2mm}
This work introduced an OT based method for interpolating the ISPCs associated with a new physical source position. By exploiting the distance invariance of corresponding image sources under source displacement as ground cost of OT, the proposed framework estimated the association and interpolation variables jointly. The numerical results demonstrate that the proposed approach can recover new ISPC in the presence of spatial perturbations and unknown associations robustly. Future work will address complex environments, nonuniform image source weights, missing or spurious image sources, computationally efficient regularized transport solvers, and validation using measured RIRs.

\vfill\pagebreak

\bibliographystyle{IEEEbib}
\bibliography{strings,refs}

@inproceedings{geldert2023interpolation,
  title={Interpolation of spatial room impulse responses using partial optimal transport},
  author={A. Geldert and N. Meyer-Kahlen and S. J. Schlecht},
  booktitle={ICASSP 2023-2023 IEEE International Conference on Acoustics, Speech and Signal Processing (ICASSP)},
  pages={1--5},
  year={2023},
  organization={IEEE}
}

@article{sundstrom2024optimal,
  title={Optimal transport based impulse response interpolation in the presence of calibration errors},
  author={D. Sundstr{\"o}m and F. Elvander and A. Jakobsson},
  journal={IEEE Transactions on Signal Processing},
  volume={72},
  pages={1548--1559},
  year={2024},
  publisher={IEEE}
}

@inproceedings{pallewela2025room,
  title={Room impulse response estimation through optimal mass transport barycenters},
  author={R. Pallewela and Y. Liu and F. Elvander},
  booktitle={2025 33rd European Signal Processing Conference (EUSIPCO)},
  pages={181--185},
  year={2025},
  organization={IEEE}
}

@article{lindheim2023simple,
  title={Simple approximative algorithms for free-support Wasserstein barycenters},
  author={J. von Lindheim},
  journal={Computational Optimization and Applications},
  volume={85},
  number={1},
  pages={213--246},
  year={2023},
  publisher={Springer}
}

@ARTICLE{KoyamaNoiseCancel,
  author={S. Koyama and J. Brunnström and H. Ito and N. Ueno and H. Saruwatari},
  journal={IEEE/ACM Transactions on Audio, Speech, and Language Processing}, 
  title={Spatial Active Noise Control Based on Kernel Interpolation of Sound Field}, 
  year={2021},
  volume={29},
  number={},
  pages={3052-3063},
  doi={10.1109/TASLP.2021.3107983}}

@ARTICLE{WaterschootDereverb,
  author={A. Jukić and T. van Waterschoot  and T. Gerkmann and S. Doclo},
  journal={IEEE/ACM Transactions on Audio, Speech, and Language Processing}, 
  title={Multi-Channel Linear Prediction-Based Speech Dereverberation With Sparse Priors}, 
  year={2015},
  volume={23},
  number={9},
  pages={1509-1520},
  doi={10.1109/TASLP.2015.2438549}}

@ARTICLE{2010SpeechEnhancement,
  author={T. Nakatani and T. Yoshioka and  K. Kinoshita and M. Miyoshi and B.-H. Juang},
  journal={IEEE Transactions on Audio, Speech, and Language Processing}, 
  title={Speech Dereverberation Based on Variance-Normalized Delayed Linear Prediction}, 
  year={2010},
  volume={18},
  number={7},
  pages={1717-1731},
  doi={10.1109/TASL.2010.2052251}}

@article{tervo2013spatialAuralization,
  title={Spatial decomposition method for room impulse responses},
  author={S. Tervo and J. P{\"a}tynen and A. Kuusinen and T. Lokki},
  journal={J. Audio Eng. Soc},
  volume={61},
  number={1/2},
  pages={17--28},
  year={2013}
}

@article{sprunck2022gridless,
  title={Gridless 3d recovery of image sources from room impulse responses},
  author={T. Sprunck and A. Deleforge and Y. Privat and C. Foy},
  journal={IEEE signal processing letters},
  volume={29},
  pages={2427--2431},
  year={2022},
  publisher={IEEE}
}

@ARTICLE{Ribeiro2012roomModeling,
  author={F. Ribeiro and D. Florencio and D. Ba and C. Zhang},
  journal={IEEE Transactions on Audio, Speech, and Language Processing}, 
  title={Geometrically Constrained Room Modeling With Compact Microphone Arrays}, 
  year={2012},
  volume={20},
  number={5},
  pages={1449-1460},
  doi={10.1109/TASL.2011.2180897}}

@article{puomio2021locating,
  title={Locating image sources from multiple spatial room impulse responses},
  author={O. Puomio and N. Meyer-Kahlen and T. Lokki},
  journal={Applied Sciences},
  volume={11},
  number={6},
  pages={2485},
  year={2021},
  publisher={MDPI}
}

@article{carlo2021dechorate,
  title={dEchorate: a calibrated room impulse response dataset for echo-aware signal processing},
  author={D. Di Carlo and P. Tandeitnik and C. Foy and N. Bertin and A. Deleforge and S. Gannot},
  journal={EURASIP Journal on Audio, Speech, and Music Processing},
  volume={2021},
  number={1},
  pages={39},
  year={2021},
  publisher={Springer}
}

@book{kuttruff2016room,
  title={Room acoustics},
  author={Kuttruff, H.},
  year={2016},
  publisher={Crc Press}
}

@article{allen1979image,
  title={Image method for efficiently simulating small-room acoustics},
  author={J. B. Allen and D. A. Berkley},
  journal={The Journal of the Acoustical Society of America},
  volume={65},
  number={4},
  pages={943--950},
  year={1979},
  publisher={Acoustical Society of America}
}

@book{pierce2019acoustics,
  title={Acoustics: an introduction to its physical principles and applications},
  author={A. D. Pierce},
  year={2019},
  publisher={Springer}
}

@article{cox2006tutorial,
  title={A tutorial on scattering and diffusion coefficients for room acoustic surfaces},
  author={T. J. Cox and B. Dalenback and P. D. Antonio and J.-J. Embrechts and J. Y. Jeon and E. Mommertz and M. Vorlander },
  journal={Acta Acustica united with ACUSTICA},
  volume={92},
  number={1},
  pages={1},
  year={2006},
  publisher={S HIRZEL VERLAG GMBH AND CO}
}

@article{ochmann2004complex,
  title={The complex equivalent source method for sound propagation over an impedance plane},
  author={M. Ochmann},
  journal={The Journal of the Acoustical Society of America},
  volume={116},
  number={6},
  pages={3304--3311},
  year={2004},
  publisher={Acoustical Society of America}
}

@article{sprunck2025fully,
  title={Fully reversing the shoebox image source method: From impulse responses to room parameters},
  author={T. Sprunck and A. Deleforge and Y. Privat and C. Foy },
  journal={IEEE Transactions on Audio, Speech and Language Processing},
  volume={33},
  pages={1023--1033},
  year={2025},
  publisher={IEEE}
}

@ARTICLE{4323129,
  author={R. W. Sittler },
  journal={IEEE Transactions on Military Electronics}, 
  title={An Optimal Data Association Problem in Surveillance Theory}, 
  year={1964},
  volume={8},
  number={2},
  pages={125-139},
  doi={10.1109/TME.1964.4323129}}

@article{BarShalom1975TrackingIA,
  title={Tracking in a cluttered environment with probabilistic data association},
  author={Y. Bar-Shalom and E. T. S. Tse},
  journal={Autom.},
  year={1975},
  volume={11},
  pages={451-460},
  url={https://api.semanticscholar.org/CorpusID:121006529}
}

@INPROCEEDINGS{7953335,
  author={Y. Guo and H. Zhu and Q. Cheng},
  booktitle={2017 IEEE International Conference on Acoustics, Speech and Signal Processing (ICASSP)}, 
  title={Indoor multi-sound source localization based on nonparametric Bayesian clustering}, 
  year={2017},
  volume={},
  number={},
  pages={6135-6139},
  doi={10.1109/ICASSP.2017.7953335}}

@INPROCEEDINGS{8455616,
  author={X. Dang and H. Zhu and Q. Cheng},
  booktitle={2018 21st International Conference on Information Fusion (FUSION)}, 
  title={Multiple Sound Source Localization Based on a Multi-Dimensional Assignment Model}, 
  year={2018},
  volume={},
  number={},
  pages={1732-1737},
  doi={10.23919/ICIF.2018.8455616}}

@INPROCEEDINGS{7077851,
  author={G. Kearney and C. Masterson and S. Adams and F. Boland },
  booktitle={2009 17th European Signal Processing Conference}, 
  title={Dynamic Time Warping for acoustic response interpolation: Possibilities and limitations}, 
  year={2009},
  volume={},
  number={},
  pages={705-709},
  doi={}}

@book{villani2009optimal,
  title        = {Optimal Transport: Old and New},
  author       = {C. Villani},
  series       = {Grundlehren der mathematischen Wissenschaften},
  volume       = {338},
  year         = {2009},
  publisher    = {Springer},
  address      = {Berlin, Heidelberg},
}

@article{peyre2020computationaloptimaltransport,
  title={Computational Optimal Transport},
  author={P. Gabriel and C. Marco},
  journal={Found. Trends Mach. Learn.},
  year={2018},
  volume={11},
  pages={355-607},
  url={https://api.semanticscholar.org/CorpusID:73725148}
}

@inproceedings{lin2020fixed,
  title     = {Fixed-Support Wasserstein Barycenters:
               Computational Hardness and Fast Algorithm},
  author    = {T. Lin and N. Ho and X. Chen and
               M. Cuturi and M. I. Jordan },
  booktitle = {Advances in Neural Information Processing Systems},
  volume    = {33},
  year      = {2020}
}

@inproceedings{cuturi2014fast,
  title     = {Fast Computation of Wasserstein Barycenters},
  author    = {M. Cuturi and A. Doucet},
  booktitle = {Proceedings of the 31st International Conference
               on Machine Learning},
  series    = {Proceedings of Machine Learning Research},
  volume    = {32},
  number    = {2},
  pages     = {685--693},
  year      = {2014}
}

@article{larsson2025single,
  title={Single-source localization as an eigenvalue problem},
  author={M. Larsson and V. Larsson and K. {\AA}str{\"o}m and M. Oskarsson},
  journal={IEEE Transactions on Signal Processing},
  volume={73},
  pages={574--583},
  year={2025},
  publisher={IEEE}
}

@inproceedings{kelley2024rir,
  author    = {L. Kelley and D. Di Carlo and A. A. Nugraha and M. Fontaine and Y. Bando and K. Yoshii},
  title     = {{RIR}-in-a-Box: Estimating Room Acoustics from {3D} Mesh Data through Shoebox Approximation},
  booktitle = {Interspeech 2024},
  pages     = {3255--3259},
  year      = {2024},
  doi       = {10.21437/Interspeech.2024-2053}
}

@article{dokmanic2013acoustic,
  title={Acoustic echoes reveal room shape},
  author={I. Dokmani{\'c} and R. Parhizkar and A. Walther and Y. M. Lu and M. Vetterli},
  journal={Proceedings of the National Academy of Sciences},
  volume={110},
  number={30},
  pages={12186--12191},
  year={2013},
  publisher={National Academy of Sciences}
}

\end{document}